\documentclass[runningheads]{llncs}
\usepackage[T1]{fontenc}

\usepackage{graphicx}
\usepackage{cite}
\usepackage{lineno}
\usepackage{subcaption}
\usepackage{multicol}
\usepackage{lipsum}
\usepackage[table,xcdraw]{xcolor}

\usepackage{algorithmic}
\usepackage{textcomp}
\usepackage{array,longtable}
\usepackage{booktabs,makecell,tabularx}
\usepackage{colortbl}
\usepackage{multirow}
\usepackage{xcolor}
\usepackage{xspace}

\usepackage{fancyhdr}
\usepackage{enumerate}

\begin{document}
\title{An Empirical Assessment of Security and Privacy Risks of Web-based Chatbots}
%
%
\author{Nazar Waheed\inst{1}\and
Muhammad Ikram\inst{2} \and
Saad Sajid Hashmi\inst{3}
\and
Xiangjian He\inst{1}
\and
Priyadarsi Nanda\inst{1}}
\author{Nazar Waheed\inst{1}
\and
Muhammad Ikram\inst{2}
\and
Saad Sajid Hashmi\inst{3}
\and
Xiangjian He\inst{1}
\and
Priyadarsi Nanda\inst{1}
}

\authorrunning{N. Waheed et al.}

\institute{University of Technology Sydney, NSW 2007, Australia 
\and
Macquarie University, NSW, Australia
\and
University of Wollongong, Wollongong, Australia
}
\maketitle              
\begin{abstract}

Web-based chatbots provide website owners with the benefits of increased sales, immediate response to their customers, and insight into customer behaviour. While Web-based chatbots are getting popular, they have not received much scrutiny from security researchers. The benefits to owners come at the cost of users’ privacy and security. Vulnerabilities, such as tracking cookies and third-party domains, can be hidden in the chatbot’s iFrame script. This paper presents a large-scale analysis of five Web-based chatbots among the top 1-million Alexa websites. Through our crawler tool, we identify the presence of chatbots in these 1-million websites. We discover that 13,515 out of the top 1- million Alexa websites (1.59\%) use one of the five analysed chatbots. Our analysis reveals that the top 300k Alexa ranking websites are dominated by {\tt Intercom} chatbots that embed the least number of third-party domains. {\tt LiveChat} chatbots dominate the remaining websites and embed the highest samples of third-party domains. We also find that 850 (6.29\%) of the chatbots use insecure protocols to transfer users’ chats in plain text. Furthermore, some chatbots heavily rely on cookies for tracking and advertisement purposes. More than two-thirds (68.92\%) of the identified cookies in chatbot iFrames are used for ads and tracking users. Our results show that, despite the promises for privacy, security, and anonymity given by the majority of the websites, millions of users may unknowingly be subject to poor security guarantees by chatbot service providers.

\end{abstract}
\section{Introduction}
	
	A Web-based chatbot (or bot) is a computer program interacting with users via a conversational user interface that simulates a conversation with a human user via textual methods~\cite{Shawar2007ChatbotsAT}.  
	Web-based chatbots offer improved customer services and efficiently manage human resources~\cite{Michiels2017ModellingCW, Ivanov2017}. For example, 
	a website owner performs customer acquisition tasks (such as new customer query, or after-sales services) through customer service (or sales and marketing) personnel. As the business gets bigger and busier, the traditional way of interacting with the online customers gets choked up resulting in increased waiting queue. Besides, the customer service representative may not be available around the clock.  
	Web-based chatbot provides a website owner with the benefits of increased sales, immediate response to their customers' queries and insights into customers' behaviours. While Web-based chatbots are getting popular, they have not received much scrutiny from security researchers. The benefits of chatbots can come at the cost of privacy and security threats. These threats are inherited by third-party domains and cookies, which might be built-in to the script. These domains and cookies can be used for the purpose of tracking users and providing personalised advertisements. There has been plethora of work done based on the security and privacy issues of a complete website~\cite{Ikram_Trackers_2016, Masood2018, hashmi2019longitudinal}. However, as per our knowledge, there is no research study that focuses explicitly on the privacy and security issues of Web-based chatbots.  

	While Web-based chatbots are getting popular and they come with several above-mentioned benefits, their advantages are inherited with several disadvantages. {\it Firstly}, consumers are concerned about their privacy and security~\cite{Michiels2017ModellingCW}. Despite the remarkable improvements in Web-based chatbots being able to mimic a human conversation, they are vulnerable to the Reconnaissance and Man-in-the-Middle attacks \cite{Carter2019}.  
	{\it Secondly}, since the chatbot is a computer program, it does not have its own identity or emotions like a real human. Customers often tend to make connections during conversations, which is lacking when engaging with chatbots. The lack of personality in chatbots and their inability to make an emotional connection is a concern for some customers. 
	{\it Finally}, a Web-based chatbot is still in its infancy since natural language processing is not the core competency in chatbot applications and is still in the development phase \cite{Michiels2017ModellingCW}.
	Web-based chatbots are prone to common communication errors, therefore, companies and organisations are very careful in using them to avoid any brand damage.

	Although several studies have taken place to study chatbots in general, none of them covers their security and privacy comprehensively. There has been extensive research on the security and privacy issues of websites, however, to the best of our knowledge, we did not find any study that focuses on the iFrame component of the Web-based chatbot for the same issues.
	
	An overview of our methodology is presented in Figure \ref{fig:overview}. 
	In this paper, we present our methodology (depicted in Figure~\ref{fig:overview}) to analyse Web-based chatbots at scale. We begin by inspecting how to filter chatbot websites by manually analysing the Alexa top 100 websites. We develop a Selenium-based web crawler to automatically detect these websites based on our analysis and assure the accuracy is 100\%. We also search for popular chatbots on the internet and select them based on their prominence. We find a total of 13,515 chatbot websites for our five selected chatbots as our dataset. 
	
	We then inspect 10 different categories of websites in our dataset. 
	We observe that Web-based chatbots present predominantly in the non-IT business category (21.78\%), IT category (16.16\%), and shopping category (5.89\%). The complete list can be seen in the Figure \ref{fig:ChatbotCategory}.
	We confirm that at least 4.2\% of the Alexa top 500k Web-based chatbots, 14.88\% of the second half of the Alexa top 1-million, and at least 6.29\% in the top 1-million Alexa Web-based chatbots are still using insecure HTTP. Although seemingly small, the fact that these are the most popular websites is a big security concern. We then proceed to inspect the Web-based chatbot iFrame in particular instead of entire website DOM to find the vulnerabilities in our selected chatbots. We find that among the three chatbots that agree to write cookies on a customer's visit to their website, {\tt Drift} chatbot writes nine different types of cookies 5,396 times out of which at least 94.62\% are tracking cookies. {\tt Hubspot} chatbot writes twelve different types of cookies 15,829 times out of which 79.35\% are tracking cookies, and {\tt Intercom} chatbot writes fourteen different types of cookies 18,995 times out of which 34.85\% are tracking and analytics cookies while 18.07\% are Ads and marketing cookies. We cannot find any cookies for {\tt Tidio} and {\tt LiveChat} chatbots, and their support team confirms this as well.  
	Note that we do not take the cookies of entire websites into account, rather we  
	focus on the cookies that a chatbot is used for essential and tracking/advertisement purposes.   
	Our focus is on the chatbot and its {\tt iFrame} only, which, to the best of our knowledge, has not been discussed in any study thus far.
	
	Despite the assurances for privacy, security, and anonymity given by the websites and privacy policies, users are victims of personally identifiable information (PII) leakages~\cite{hashmi2021compliance}. Similarly, by using chatbot services, users may inadvertently be exposed to the privacy and security risks~\cite{Masood2018}.   
	
	In summary, the contributions of this paper are as follows.
	
	\begin{enumerate}
	    \item We present the first large-scale study of security and privacy issues in chatbots on Alexa top 1-million popular websites~\cite{Amazon}. 
	    We detect 13,515 (1.59\%) websites leveraging web chatbots for customers' interaction. We release our data and scripts for future research. 
	   
	    \item We analyse the 13,515 (1.59\%) websites for the type of chatbots and analyse the coverage of the detected chatbots. We find that  
	    21.78\% of the chatbot websites belong to the non-IT business category, while the percentage of Information Technology (IT) chatbot websites is 16.16\%, and shopping with 5.89\% is the third most dominant category. We also analyse the security and privacy issues of our dataset chatbot websites. We explore the chatbot websites and find that 6.29\% of them are still using the insecure HTTP protocol, where an alarming 14.88\% of the websites ranking >500k still transfer their visitors' data in plaint-text. This shows that among the most popular websites, non-IT business, IT and shopping websites are more vulnerable than any other website categories. 
	   
	    \item Our analysis illuminates that chatbots have a disproportionate use of cookies for tracking and \emph{essential} or \emph{useful} functionalities. We discover 5,396 cookies in 2,110 websites leveraging {\tt Drift} chatbot. 5,113 (94.62\%) and 283 (5.24\%) of the cookies are used for Tracking and essential functionalities, respectively. 
	    On the other hand, 2,185 websites rely on {\tt Hubspot} for the provision of chat services via a total number of 15,829 unique cookies with 79.35\% (12561) for tracking while the rest are essential cookies.  
	    
	   
	   \item We identify the top 10 third-party domains embedded in the iFrames of each web-based chatbot. The most common third-parties are well-known operators, for example, googleapis, cloudflare, w3, and facebook. These operators are imported by 39.67\% (5361), 15.43\% (2085), 6.1\% (822), and 3.35\% (453) web-based chatbots, respectively.  
	   
	    \end{enumerate}
The rest of this paper is organized as follows: In Section \ref{sec:ConceptsTerms}, we present concepts and terms related to web tracking and services. We present our methodology for web-based chatbot detection and data collection in Section~\ref{sec:ChatbotDetectionMethod}. In Section~\ref{sec:Exploring}, we analyse our chatbots in the top 1-million Alexa websites and present our findings such as the presence of chatbots on websites, tracking cookies, and third-party domains. Section~\ref{sec:RelatedWork} presents the related work while we conclude our work by presenting the gaps with some future directions in Section~\ref{sec:Conclusion}.

\section{Concepts and Terms}
\label{sec:ConceptsTerms}
We begin by introducing the general concepts and terms used in the paper.  

\textbf{Advertising and tracking domain:} The \textit{advertising and tracking} domain (or tracker) is the URL of an entity embedded in a web page. The purpose of a tracker is to re-identify a user's visit on the web page again for loading custom themes or analytics ({\it first-party} tracking) or to re-identify a given user across different websites for building the user's browsing profile or providing personalised advertisements ({\it third-party} tracking).

\textbf{iFrame:} An iFrame or inline Frame is an HTML document embedded within an HTML web page. The purpose of an iFrame is to display embedded HTML contents from a different web page into the current web page. The contents of iFrames can be videos, maps, advertisements, chatbot services, as well as tracking components like cookies and JavaScript codes. Hence, besides providing utilities and services, iFrames can also be used for third-party tracking.

\textbf{Cookie:} A cookie (or HTTP cookie) is a text file that is stored on the user's device by the web browser. The content of a cookie is in plain text format. A cookie is generated by the web server (of a web page) and is sent back from the user's device to the web server at each subsequent visit by the user. A cookie can store shopping carts, theme preferences of the user, or the user's authentication status. Cookies generated by third-parties via iFrames can be used for third-party tracking. Different types of cookies are discussed in detail in Section~\ref{subsec:AnalysisCookies}.

\section{Chatbot Detection Methodology and Dataset}
\label{sec:ChatbotDetectionMethod}
\begin{figure}[ht]
\centering
\includegraphics[width=0.8\textwidth]{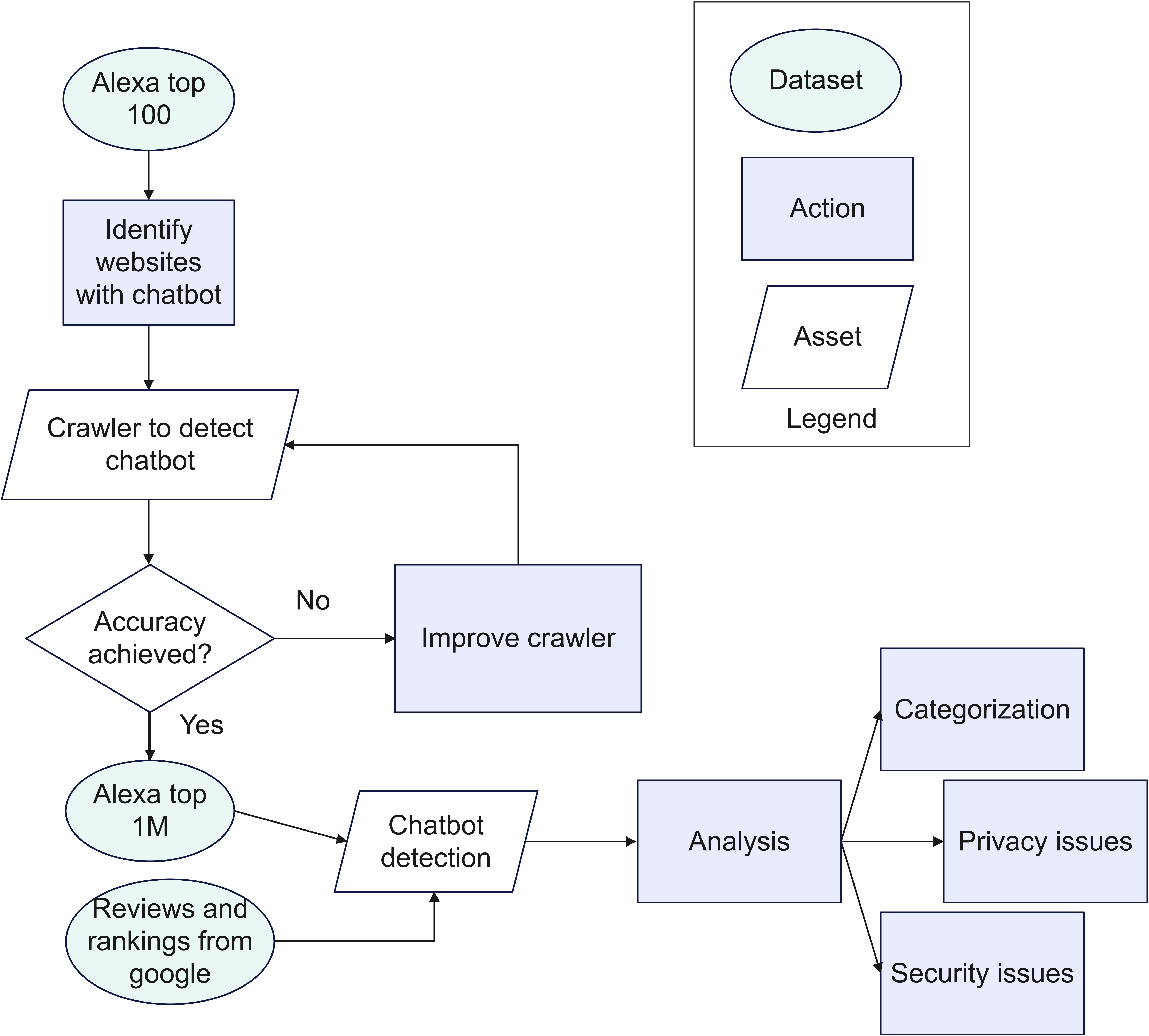}
\caption{\small Overview of our crawling and analysis methodology: We manually inspect the top 100 Alexa websites for chatbots to identify chatbot services and to construct keywords list for automatic detection of chatbots in the top 1-million Alexa websites. We then perform an analysis to categorise websites and to analyse security and privacy issues.}
\label{fig:overview}
\end{figure}

We begin by presenting our methodology, over-viewed in Figure \ref{fig:overview}, for detecting chatbots employed in the top 1-million Alexa websites. We then characterise our dataset. 

\subsection{Discovering Chatbots} 
Using Selenium Web Driver, we develop an automated web crawler to automate the visiting and rendering process of analysed websites. To increase our chatbot coverage and maximise the number of detected chatbots, we implement a crawler framework. We begin by discovering web-based chatbot services on the Alexa top 1-million websites. To this end, we find the difference between a normal website and the website with a chatbot service. We manually inspect the first Alexa top 100 websites for \emph{potential} web chatbot services. Typically, websites implement chatbot services in iFrames, therefore, we explicitly focused on the iFrame of the chatbot on these 100 websites. The keywords include: 
`\textit{chat widget}', \textit{let's chat}, \textit{drift-widget}, `\textit{chat now}', and `\textit{chatbot}'. While we acknowledge that our keywords list is not exhaustive to include chatbots on non-English language websites, we do consider our method for chatbots as a \emph{lower-limit} on the number of chatbots on the top 1-million Alexa websites. 

 Next, we crawl through the chatbot websites and extract their chatbot iFrame cookies only instead of the whole website, since we are specifically interested in the security and privacy issues of the web-based chatbots.  We then analyse the embedded third-party domains in each of those chatbots. To extract the third-party domains, we only check the contents of the iFrame of a chatbot, instead of the complete website's DOM. Overall, we find 13,515 (1.6\%) chatbot websites, out of which 566 (4.2\%) websites do not render, either due to the website being closed or moved permanently to a new domain name.

{\bf Issues and Limitations.} For chatbot websites, once a website is completely rendered, the chatbot icon is found at the bottom right corner of the screen. Sometimes, the chatbot is not visible on the respective website. This is mostly due to one of the following reasons  ({\it i}) the chatbot is only available during certain office hours, and ({\it ii}) the chatbot is offline/hidden as the developers may be working on it. 
    
\subsection{Data Augmentation} 
\begin{figure}[ht]
  \centering
  \includegraphics[width=0.8\linewidth]{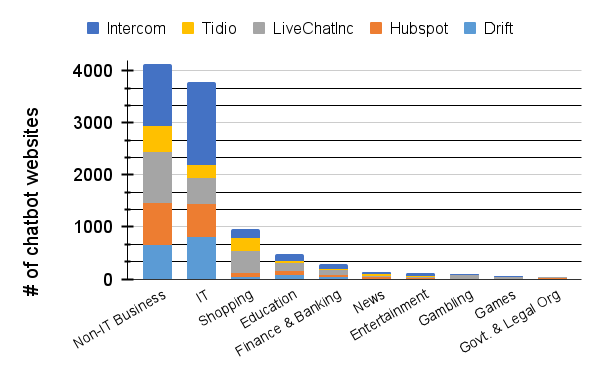}
  \caption{Categories of chatbot websites in the top 1-million Alexa websites. }
  \label{fig:ChatbotCategory}
\end{figure}

Next, to analyse the coverage of chatbots in various categories of websites, we aim to categorise the Alexa top websites. There are several databases and tools available and website categories stored. However, we use crawling techniques on \textit{Fortiguard} website classification tool \cite{Fortiguard} 
to gather this information. The websites that return errors while rendering in the first phase are manually labelled. We find the categories of each chatbot website in our dataset (13,515 websites). The top 10 categories, depicted in Figure \ref{fig:ChatbotCategory}, are selected based on their frequencies of occurrence. These ten categories comprise 75\% of our dataset, while the remaining 25\% are categorised as \textit{Others}.  It is found that most of the chatbot websites are used by \textit{non-IT Business} and \textit{IT} category websites. However, chatbots are not a popular choice among \textit{Games} and \textit{Government and Legal Organizations} related website owners. 

\begin{table}[ht]
    \resizebox{\columnwidth}{!}{%
    \centering
    \begin{tabular}{ c|c|c|c|c|c|c  }
    \toprule\toprule
    & \multicolumn{5}{|c|}{\bf Chatbots} & { \# Total (\%)} \\
    \midrule
    {\bf Alexa Rank} & {\bf Drift} & {\bf LiveChat } & {\bf Hubspot} &{\bf Tidio} & {\bf Intercom}  \\
    \midrule
    1-100K & 619 (0.62\%) & 436 (0.44\%) & 210 (0.21\%) & 100 (0.10\%) & \cellcolor[HTML]{DAF7A3}940 (0.94\%) & 2,305 (2.31\%) \\
    100K-200K & 453 (0.45\%) & 591 (0.59\%) & 386 (0.39\%) & 185 (0.19\%) & \cellcolor[HTML]{DAF7A3}757 (0.76\%) & 2,372 (2.37\%) \\
    200K-300K & 407 (0.41\%) & 586 (0.59\%) & 394 (0.39\%) & 343 (0.34\%) & \cellcolor[HTML]{DAF7A3}681 (0.68\%) & 2,411 (2.41\%) \\
    300K-400K & 295 (0.30\%) & \cellcolor[HTML]{DAF7A3}573 (0.57\%) & 413 (0.41\%) & 365 (0.37\%) & 531 (0.53\%) & 2,177 (2.18\%) \\
    400K-500K & 150 (0.15\%) & \cellcolor[HTML]{DAF7A3}433 (0.43\%) & 305 (0.31\%) & 286 (0.29\%) & 429 (0.43\%) & 1,603 (1.60\%)\\
    500K-600K & 65 (0.07\%) & \cellcolor[HTML]{DAF7A3}319 (0.32\%) & 160 (0.16\%) & 218 (0.29\%) & 273 (0.27\%) & 1,035 (1.04\%)\\
    600K-700K & 51 (0.05\%) & \cellcolor[HTML]{DAF7A3}341 (0.34\%) & 143 (0.14\%) & 125 (0.13\%) & 182 (0.19\%) & 843 (0.84\%) \\
    700K-800K & 36 (0.04\%) & 131 (0.13\%) & 64 (0.06\%) & 2 (0.002\%) & \cellcolor[HTML]{DAF7A3}136 (0.14\%) & 369  (0.37\%)\\
    $\geq$ 800K & 26 (0.05\%) & 86 (0.18\%) & \cellcolor[HTML]{DAF7A3}104 (0.22\%) & 50 (0.11\%) & 93 (0.20\%) & 359 (0.75\%)\\
    \midrule
    {\bf Overall (1-million)} & 2,110 (0.25\%) & 3,507 (0.41\%) & 2,185 (0.26\%) & 1,676 (0.20\%) & \cellcolor[HTML]{DAF7A3}4,037 (0.48\%) & {\bf 13,515 (1.59\%)}\\
    \bottomrule\bottomrule
    \end{tabular}}
    \vspace{0.1cm}
    \caption{\small Frequency (and percentage) of chatbot services amongst the Alexa top 1-million websites. Highlighted trends show {\tt Intercom} chatbot is the preferred choice for the most popular set of websites followed by {\tt LiveChat} which is also the preferred choice for the next tier of popular websites.} 
    \label{tab:ChatbotWebsites}
    
\end{table}

\subsection{Dataset}
Our comprehensive analysis is done by breaking the dataset into parts with each part having 10,000 websites to get an in-depth measurement of our study. Based upon the keywords (cf. \S~\ref{sec:ChatbotDetectionMethod}), we run our crawler that detect chatbots on 3.5\% of the analysed websites. To check the accuracy of our crawler, we manually label the first hundred Alexa ranking websites and perform manual testing on them. It is learnt that our model is ~61\% accurate. The reason is that there are several possible ways to write a website script, and using the keywords alone is not an optimum solution. 
    
\begin{figure}
  \centering
  \includegraphics[width=0.8\linewidth]{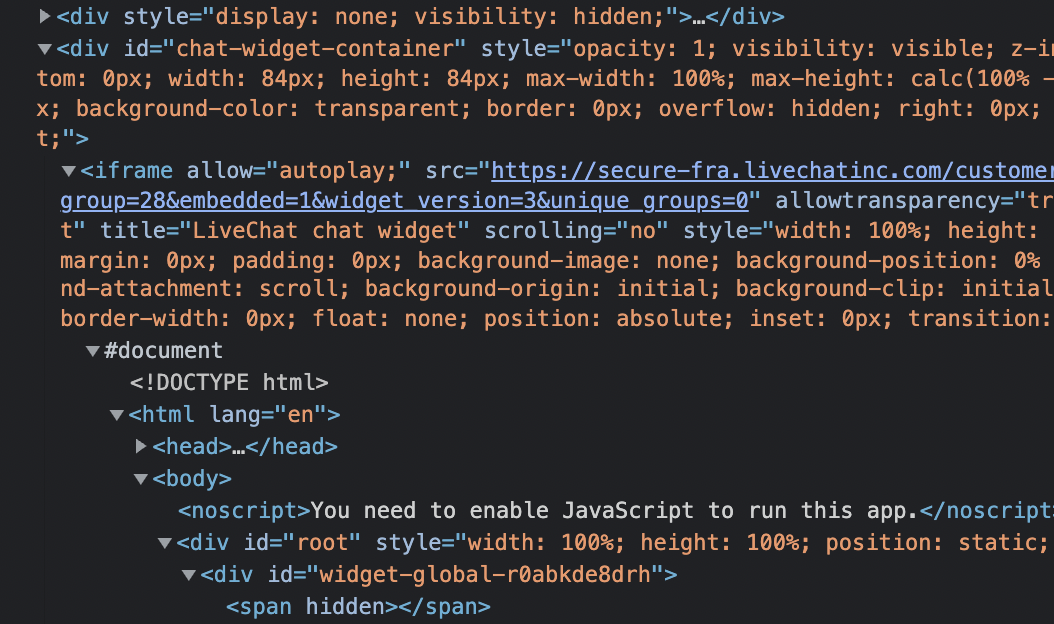}
  \caption{An example of an iFrame enabling a typical chatbot service on a given website.}
  \label{fig:iFrame}
\end{figure}

Finding a common script, or tag among all of them is not possible. However, we find some unique keywords/tags/elements. Figure \ref{fig:iFrame} shows the iFrame of a chatbot website \url{www.synology.com}. It has a tag {\itshape id='chat-widget-container'}, which can be used to filter the {\tt LiveChat} chatbot websites. Similarly, we select five chatbots: {\tt LiveChat}, {\tt Drift}, {\tt Intercom}, {\tt Tidio} and {\tt Hubspot} based on their frequencies of occurrence in the top 10k Alexa ranking websites. Overall, our crawler identifies 13,515 chatbot websites from Alexa top 1-million websites. 

Table \ref{tab:ChatbotWebsites} summarises our findings. We observe that the
{\tt Intercom} chatbot is the preferred choice for the most popular set of websites (top 300k) followed by {\tt LiveChat} for the next tier of Alexa ranking websites. Overall, {\tt Intercom} chatbot is found on 29.87\% of them, {\tt LiveChat} on 25.95\%, {\tt Drift} on 15.61\%, {\tt Hubspot} on 16.17\%, and {\tt Tidio} on 12.40\% only. 
    
Based on the above findings, in the first round, we crawl the top 10k websites and render their DOMs. After optimising our crawler, we can filter all chatbots with 100\% accuracy. 
    
We also search for the top Web-based chatbots by using different keywords over the google search. We find chatbot rankings and reviews on the websites in \cite{IndhujaLal, AaronBrooks, WernerGeyser, Leah, ShaneBarker, SyedBalkhi, Zfort}  (accessed in Feb 2022). We choose the top three chatbots. After selecting \textit{MobileMonkey}, \textit{Aivo}, and \textit{Pandorabots} from the blogs and reviews, we run our automated scraper for the top 200k websites and find only two chatbots belonging to \textit{MobileMonkey}, four chatbots to \textit{Botsify}, and zero for both \textit{Aivo} and \textit{Pandorabots}. Therefore, due to their insignificant presence, we do not consider them in our analysis further.
    
As a second attempt, we manually re-analyse the top 100 websites and find two relevant chatbots (\textit{SF-chat} and \textit{SnatchBot}) and search for them over the top 10k websites using an automated script. For \textit{Salesforce} chatbot, we only find it to be on their own websites, for example, \textit{cloudforce} and \textit{exactforce}. On all other top 10k websites, we do not find any other websites having either of these chatbots. 729 websites do not render in the first phase, and they are analysed again in the second attempt (we learn that rendering chatbot websites take longer than our previous timeout). We also manually analyse the 100 chatbots from 100,000 to 100,100 range and find three chatbots only, i.e., (i) {\tt Drift}, (ii) {\tt Intercom}, and (iii) {\tt eLum}\footnote{\url{https://eluminoustechnologies.com}} . {\tt Drift} is already included in our study, {\tt Intercom} is found on numerous websites (after initial automated crawler verification), and {\tt eLum} is not found anywhere else since it is a private custom chatbot. Moreover, please note that social-media related chatbots like Facebook messenger are not valid chatbots since they require human interaction and are not automated. Therefore, we do not include them in our analysis. For the rest of the study, we use only five chatbots, which are {\tt Drift}, {\tt Hubspot}, {\tt LiveChat}, {\tt Tidio}, and {\tt Intercom}.

\section{Exploring Web Chatbots}
\label{sec:Exploring}

\subsection{Analysis of HTTP Chatbot Websites}
\label{subsec:AnalysisHTTP}

\begin{figure}
    \centering
    \subfloat[\label{a}]{\includegraphics[scale=0.16, keepaspectratio]{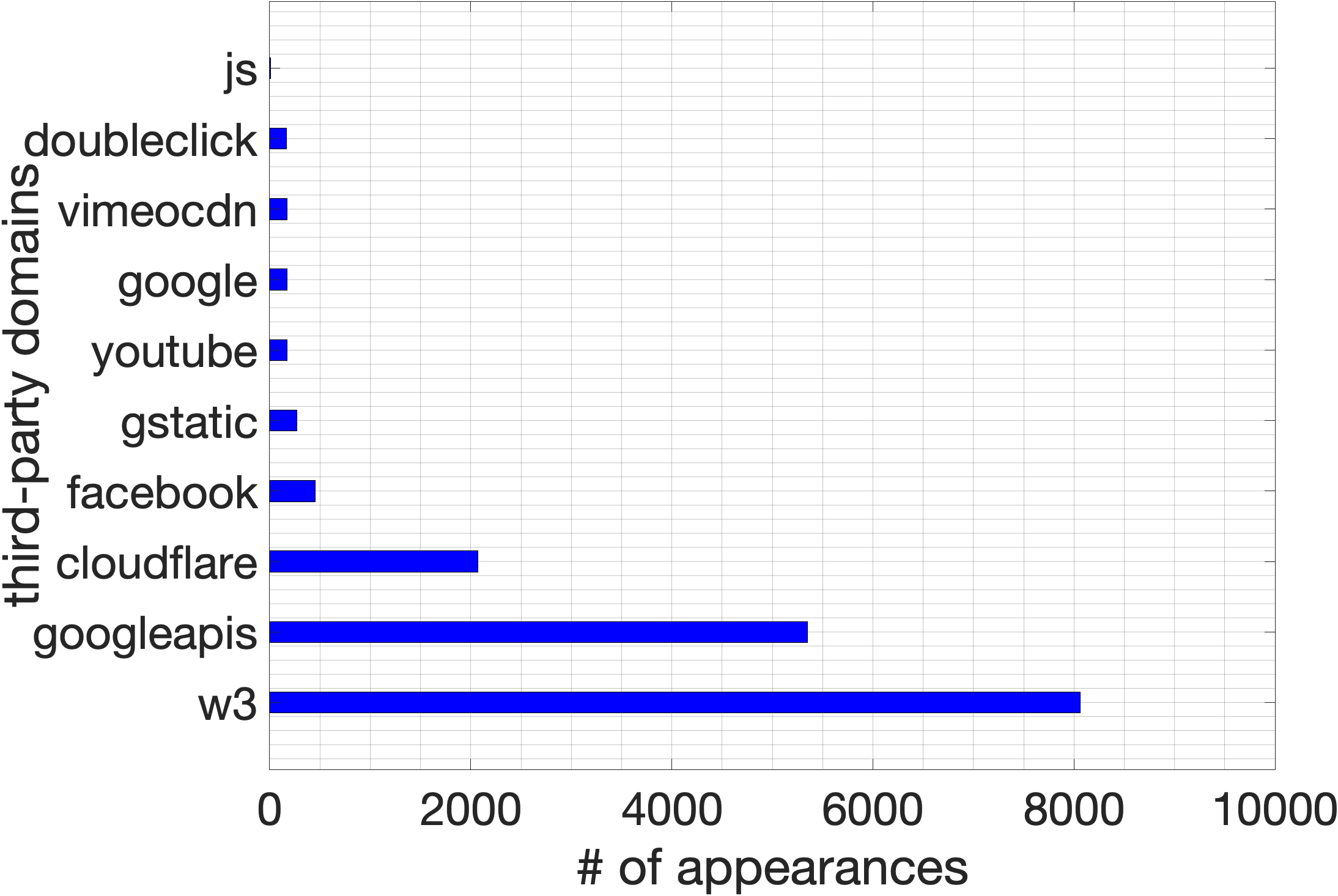}\label{fig:thirdPartyDomains}}\quad
    \subfloat[\label{b}]{\includegraphics[scale=0.16, keepaspectratio]{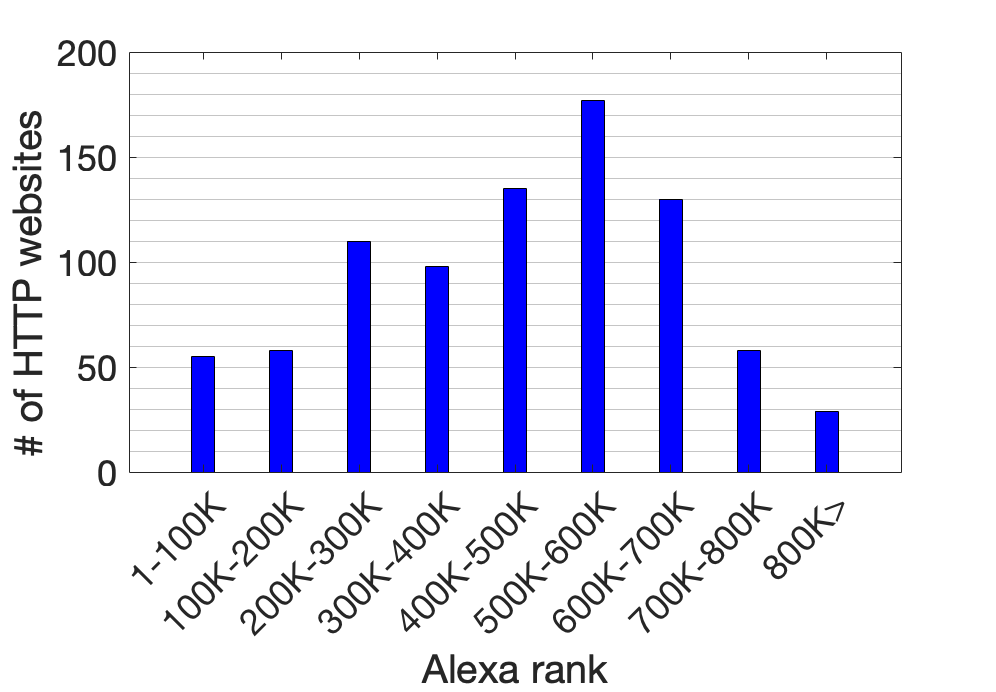}\label{fig:HTTP}}
    \caption{\small (a) Breakdown of third-parties found in web-based chatbots. (b) Number of web-based chatbots using insecure HTTP websites in top Alexa websites.}
\end{figure}

To check whether a website uses HTTP, our crawler defaults to communicating with the site over HTTP by simply concatenating the 'http://' or 'http://www.' string with the hostname provided in the Alexa data. 
Once the crawler receives a final response and it does not redirect the client requests from HTTP to HTTPs, it is marked as HTTP. 
We also check the websites that have errors by manually inspecting each one of them and discover that such websites are very few and the main reason for the errors is that they do not exist anymore (something that Alexa should take care of as it is not updated). The trend in the Figure \ref{fig:HTTP} shows that less popular websites are less secure. We find that 850 (6.29\%) out of 13,515 chatbot websites are still using the insecure HTTP version.

    \subsection{Analysis of Cookies }
    \label{subsec:AnalysisCookies}
    The online ecosystem is composed of a large number of organizations engaging in tracking user behaviour across the web~\cite{cook2019inferring}. This is accomplished by a variety of techniques including tracking cookies, pixel tags, beacons, and other sophisticated mechanisms. Below, we provide an 
    overview of the most common cookies. 
    
    \paragraph{Identification Cookies}
    These cookies can track visitor's conversations and interactions with a website. The customer service representative uses such information to offer better service. It is challenging to learn about any old chat with the customer without these cookies.
    
    
    \paragraph{Tracking Cookies}
    
    These are the most common cookies used now to track user behaviour, user information and visits to a website.
    
     \paragraph{Performance and Functionality Cookies:}
    These cookies are used to enhance the performance and functionality of a website but are non-essential to their use. However, certain functionalities like videos may become unavailable or the login details are required every time a user visits the website. 
    
    \paragraph{Conditional Cookies}
    These cookies may be written onto a website since they depend on using a specific feature of a website. 
    
    \paragraph{Marketing Cookies:}
    These are account-based marketing cookies used to identify prospects and personalise sales and marketing interactions.
    
    \paragraph{Analytics and Customization Cookies:}
    These cookies are used to determine the effectiveness of marketing campaigns. Website owners use them to collect limited data from end-user browsers to enable them to understand the use of their websites. 
    
    
    \paragraph{Advertising Cookies:}
    These cookies collect information over time about users' online activity on the websites and other online services to customise online advertisements.
    
    The details about all cookies used on every chatbot can be read on their website \cite{Drift_Cookies,Intercom, Hubspot2021,LiveChat,Tidio}.

    \subsubsection{Drift:}
    According to {\tt Drift}, the primary reason it uses cookies is to track user interactions with the visited website. It also uses cookies to customise products to the need of a customer. {\tt Drift} claims that the data is never sold or sent to third-parties. Instead, it is used in their platform to allow for more personalised and specific messaging \cite{Drift_Cookies}.

    \subsubsection{Hubspot:}
    According to {\tt Hubspot}, it uses cookies to track users who visit a {\tt Hubspot} chatbot website. The purpose of these cookies is to keep track of visit counts and information about the sessions (such as session start timestamp). When the {\tt Hubspot} software is run on a website, it leaves behind these cookies to help {\tt Hubpost} identify the users on future visits \cite{Hubspot2021}. 
    
    
    
    

    
    \subsubsection{LiveChat:}
    We search for {\tt LiveChat} cookies manually by inspecting several websites. We do not find any tracking cookie in our manual search. To confirm, we inquire from the {\tt LiveChat} support team to ensure that none of the cookies is used for tracking purposes. The support team confirmed the same. The {\tt LiveChat} chatbots automatically save and store two essential cookies on the user's device when a user visits a website with {\tt LiveChat} widget \cite{LiveChat}. The two essential cookies are as follows:
    
    \paragraph{\_\_lc\_cid (customerID)}
    This is a functional cookie that {\tt LiveChat} account service uses. The purpose of this cookie is to verify the identity of a customer created.
    
    \paragraph{\_\_lc\_cst (customerSecureToken)}
    This is also a functional cookie that {\tt LiveChat} account service uses to identify a user, for example, name, IP address, geolocation etc.
     
    \subsubsection{Tidio:}
    According to {\tt Tidio}, it uses cookies to maintain, improve and customise the user experience. Additionally, the cookies are used to remember the visitor's choice, such as language preference. {\tt Tidio} claims to collect information, including PII and assures that it will be used by them only. We cannot find any evidence of {\tt Tidio} cookies on any of the websites using their chatbots, nor can we find any information about what cookies are used on their website \cite{Tidio}.
    
    \subsubsection{Intercom:}
    According to {\tt Intercom}, its chatbot writes “first-party” cookies only and assures that its cookies are strictly private and confidential. The purpose of these cookies is to identify users and keep track of sessions. Intercom states that it uses two cookies only \cite{Intercom}; however, this claim is contradictory to our findings discussed below

\begin{table}[t]
{%
\setlength{\tabcolsep}{4pt}

\begin{tabular}{l|c|c|c|c}
\toprule\toprule
\multirow{2}{*}{} & \multicolumn{3}{c|}{\textbf{Categories of Cookies}} & \multirow{2}{*}{} \\
\cmidrule(lr){2-4}
 {\textbf{Chatbots}} & \textbf{Essential} & \textbf{Tracking \& Analytics} & \textbf{Ads \& Marketing} & \textbf{Total}  \\
\midrule
Drift & 283 (5.38\%) & 5,113 (94.62\%) & - & 5,396   \\
Hubspot & 3,268 (20.65\%) & 12,561 (79.35\%) & -  & 15,829  \\
Intercom & 8,942 (47.08\%) & 6,620 (34.85\%) & 3,433 (18.07\%) & 18,995   \\
\midrule
{\bf Total} & 12,493 (31.06\%) & 24,294 (60.4\%) & 3,433 (8.54\%) & 40,220 \\
\bottomrule\bottomrule

\end{tabular}}
\label{tab:cookies}
\vspace{0.1cm}
\caption{{\small Distribution of cookies across Web-based chatbots.}}
\end{table}

    \subsubsection{Findings/Discussion:}
    To distinguish between a first-party and a third-party cookie, we consider any cookie with the same name as the respected chatbot as a first-party. We also consider the cookies that chatbot service providers have mentioned on their websites as first-party. We declare any other cookie as a third-party.
    We find a total number of 2,110 websites using {\tt Drift}. From these, a total of 5,396 cookies are discovered. 5,113 (94.62\%) of them are used for Tracking, and 283 (5.24\%) are essential cookies.
    Hubspot is used on 2,185 websites, which have 15,829 cookies. 12,561 (79.35\%) are tracking, while the rest are \textit{essential} cookies.
    {\tt Intercom} chabot websites are 4,037, generating 18,995 cookies, out of which 52.92\% are either tracking, advertisement or marketing cookies, while 47.08\% are essential cookies for functionality. 
    No cookies are found on either {\tt LiveChat} or {\tt Tidio} chatbots. Overall, more than two thirds of the discovered cookies are used for tracking or advertisement purposes.

    \subsection{Analysis of Third-party Domains}
    \label{subsec:AnalysisTPD}
    We parse the URLs from the chatbot iFrames, extract the second-level domains using \textit{tldextract}\footnote{\url{https://pypi.org/project/tldextract/}}, and compare them with the respective website. If they match, it is declared a first-party domain; otherwise, it is stated as a third-party domain. For instance, we extract {\tt googleapis.com} and {\tt drift.com} domains from the iFrame of {\tt Drift} chatbot embedded in the landing page of \url{https://www.drift.com}. Given that {\tt googleapis.com} does not match with {\tt drift.com}, our method labelled {\tt googleapis.com} as third-party whilst {\tt drift.com} as first-party. For instance, we extract {\tt googleapis.com} and {\tt drift.com} domains from the iFrame of {\tt Drift} chatbot embedded in the landing page of \url{https://www.drift.com}. Given that {\tt googleapis.com} does not match with {\tt drift.com}, our method labelled {\tt googleapis.com} as third-party whilst {\tt drift.com} as first-party.

\begin{table}[!t] 
\resizebox{\columnwidth}{!}{%
\setlength{\tabcolsep}{4pt}
\begin{tabular}{l|c|c|c|c|c|r}
\toprule\toprule
\textbf{Third-party Domain} &
\textbf{Drift} & 
\textbf{LiveChat} &
\textbf{Intercom} & 
\textbf{Tidio} & 
\textbf{Hubspot} & \textbf{Total} \\ \midrule
w3.org	&	742	&	5	&	0	&	6,501	&	813	&	8,061	\\
googleapis.com	&	28	& 3,502	&	0	&	1,537	&	282	&	5,349	\\
cloudflare.com	& 2,063	&	1	&	0	&	0	&	10	&	2,074	\\
facebook.com	&	0	&	0	&	0	&	0	&	453	&	453	\\
gstatic.com	&	0	&	0	&	0	&	0	&	268	&	268	\\
youtube.com	&	0	& 0	&	0	&	0	&	174	&	174	\\
google.com	&	0	&	0	&	0	&	0	&	172	&	172	\\
vimeocdn.com	&	0	&	0	&	0	&	0	&	171	&	171	\\
doubleclick.net	&	0	&	0	&	0	&	0	&	166	&	166	\\
 rlets.com	&	0	&	0	&	5	&	0	&	0	&	5	\\
 {\it other domains} & 0 & 19 & 0 & 7 & 3,872 & 3,989 \\
\midrule
Total	&	2,833	&	3,527	&	5	&	8,045	&	2,053	&	{\bf 20,791}	\\

 \bottomrule\bottomrule
\end{tabular}
}
\vspace{0.1cm}
\caption{Distribution of top ten third-parties embedded in the iFrames of Web-based chatbots.}

\label{tab:3pd}
\end{table}
 
 Figure \ref{fig:thirdPartyDomains} depicts, and Table~\ref{tab:3pd} lists the top 10 third-party domains embedded in the iFrames of chatbots. We observe that all chatbots rely on third-party services such as W3, Google APIs, and CloudFlare for iFrame template, fonts, and hosting as well as storing content, respectively.  
 We observe that only one third-party domain ({\tt rlets.com}) is found on the {\tt Intercom} websites. Since {\tt Intercom} dominates the top 300k Alexa websites (52\% of total web-based chatbot websites) suggesting that the top websites do not rely much on advertising and analytical services revenues funneled from chatbots. 
 On the other hand, less popular websites generate 99.9\% of the top ten third-party domains. {\tt Hubspot} based websites have the most variety\footnote{Drift=3, Livechat=10, Hubspot=145, Tidio=4, Intercom=1} of third-party domains, making it the most vulnerable. One hundred forty five different third-party domains 
 are present in {\tt Hubspot} websites.

    
    
    


	\section{Related Work}
	\label{sec:RelatedWork}
	To the best of our knowledge, no prior work has been done to address the privacy and security risks of cookies or third-party scripts embedded in web-based chatbots. Previous work has analysed PII leaks via advertisements and third-party scripts on various domains such as Facebook~\cite{Andreou_FB_2018, Andreou_FB_2019, Venkatadri2019, Ghosh2019}, mobile eco-system~\cite{Ikram_IMC_2016, Ikram_AdBlocking_2017, hashmi2019optimization}, and web forms~\cite{Starov2015}.
		
	There are security and privacy risks associated with chatbots~\cite{Gondaliya2020}. In  \textit{financial} chatbots, Bhuiyan et al. proposed a chatbot leveraging a private blockchain platform to conduct secure and confidential financial transactions~\cite{Bhuiyan2020}. Chatbots have also been developed to remove sensitive information from the conversation before passing it to its NLP engine ~\cite{Biswas2020}. Meanwhile, threats on the chatbot's client-side (such as unintended activation attacks and access control attacks) and network-side (such as MITM attacks and DDoS attacks) have been studied in the literature~\cite{Ye2020}. Bozic et al. conducted a preliminary security study on an open-source chatbot to identify XSS and SQLi vulnerabilities~\cite{bozic2018security}. Their work did not find any XSS and SQLi vulnerabilities and was limited to analysing only one chatbot. No prior work has been done to study the iFrames of Web-based chatbots and to determine the types of cookies embedded. In this paper, to fill the gap, we study the prevalence of five chatbots in Alexa top 1-million websites and analyse the chatbot cookies and third-party domains embedded in the iFrames of chatbots.

		
	\section{Conclusion and Future Work}
	\label{sec:Conclusion}
	In this paper, 
	\emph{firstly}, we have presented the difference between websites with and without chatbots. We have found the keywords to detect chatbots on the analysed websites. We have also manually inspected the top 1,000 websites to validate chatbot detection.
    \emph{Secondly}, we have designed and implemented a crawler tool that systematically explores and collects DOMs from the top 1-million Alexa websites. We have discovered that a subset of 13,515 (1.59\%) of these websites use our five selected chatbots. We have found the frequencies of these chatbots in ten different categories and discovered that non-IT business websites had used 21.78\% of them. 
    Our analysis has revealed that the top 300k Alexa ranking websites are dominated by {\tt Intercom}, while {\tt LiveChat} dominates the remaining chatbot websites. We have also found that 6.29\% of the chatbot use insecure protocols to transfer users' chats in plain-text. 
    Our results show that, despite the promises for privacy, security, and anonymity given by the majority of the websites, millions of users may be unawarely subject to poor security guarantees by chatbot service providers on the same websites. 
    
    In the future, we want to extend our findings to the distribution of third-party domains and trackers in categories of web-based chatbots websites.  
    This will help analyse and identify the dependence of chatbot websites on advertising and analytical services. Another area to explore is whether any chatbot websites render content that it does not directly load. Informed by the study by Ikram et al. \cite{Ikram_chainOF}, this work can be extended to analyse the dependency web-resources chains of the chatbots.

\bibliographystyle{splncs04}
{
\bibliography{chatbotref}
}

\end{document}